\documentclass{article}

 \usepackage{amsmath}
\usepackage[preprint]{neurips_2025}

\usepackage[utf8]{inputenc} 
\usepackage[T1]{fontenc}    
\usepackage{hyperref}       
\usepackage{url}            
\usepackage{booktabs}       
\usepackage{amsfonts}       
\usepackage{nicefrac}       
\usepackage{microtype}      
\usepackage{graphicx}
\usepackage{xcolor,soul}         

\title{Position: Human-Robot Interaction in Embodied Intelligence Demands a Shift From Static Privacy Controls to Dynamic Learning}

\author{%
  Shuning Zhang \\
  Tsinghua University\\
  Beijing, China\\
  \texttt{zsn23@mails.tsinghua.edu.cn} \\
  \And
  Hong Jia \\
  University of Auckland \\
  Auckland, New Zealand \\
  \texttt{hong.jia@auckland.ac.nz} \\ 
  \And 
  Simin Li \\
  Beihang University \\ 
  Beijing, China \\
  \texttt{lisiminsimon@buaa.edu.cn} \\
  \And 
  Ting Dang \\
  University of Melbourne \\
  Melbourne, Australia \\
  \texttt{ting.dang@unimelb.edu.au} \\
  \And 
  Yongquan `Owen' Hu \\
  Augmented Human Lab, National University of Singapore \\
  Singapore \\
  \texttt{yongquan@ahlab.org}
  \And
  Xin Yi\thanks{Corresponding author.}\\
  Tsinghua University \\
  Beijing, China \\
  \texttt{yixin@tsinghua.edu.cn} \\
  \And 
  Hewu Li \\ 
  Tsinghua University \\
  Beijing, China \\
  \texttt{lihewu@cernet.edu.cn} \\
}

\begin{document}

\maketitle

\begin{abstract}
    The reasoning capabilities of embodied agents introduce a critical, under-explored inferential privacy challenge, where the risk of an agent generate sensitive conclusions from ambient data. This capability creates a fundamental tension between an agent's utility and user privacy, rendering traditional static controls ineffective. To address this, this position paper proposes a framework that reframes privacy as a dynamic learning problem grounded in theory of Contextual Integrity (CI). Our approach enables agents to proactively learn and adapt to individual privacy norms through interaction, outlining a research agenda to develop embodied agents that are both capable and function as trustworthy safeguards of user privacy.
\end{abstract}

\section{Introduction}

Embodied agents capable of autonomous decision-making are increasingly prevalent in human-centric environments~\cite{kimopenvla,zhen20243d}, offering personalized assistance by navigating dynamic settings~\cite{zhen20243d}, understanding nuanced commands~\cite{zhen20243d} and executing multi-step tasks~\cite{zitkovich2023rt}. The efficacy, however, depends on continuous perception and deep contextual reasoning, which creates a fundamental privacy conflict. While traditional risks~\cite{dietrich2023should,callander2024navigating,lutz2021privacy} in perception~\cite{shome2023robots}, communication~\cite{tonkin2017embodiment}, and data handling~\cite{li2024privacy} are recognized, the most critical and overlooked challenge is inferential privacy. This risk arises when an agent’s internal reasoning synthesizes multiple, non-sensitive inputs to form new, highly sensitive conclusions about a user, creating an acute tension between the agent's inferential utility and the user's privacy~\cite{yu2024panav,chatzimichali2020toward}.


Existing privacy protection strategies are illy-suited for embodied agents, particularly concerning inferential privacy. General safeguards like federated learning~\cite{miao2025fedvla} and differential privacy~\cite{zhan2025surgeon} can impair performance and be computationally costly. Furthermore, data-centric controls such as obfuscation~\cite{kim2019privacy} may destroy the contextual information essential for robust reasoning, thereby undermining the agent's ability to make accurate inferences. This creates a trade-off that measures designed to protect privacy often degrade the inferential capabilities that define an agent's utility, hindering users' adoption. Compounding this, privacy preferences are highly personal and context-dependent~\cite{asthana2024know,nissenbaum2004privacy}, yet recent studies show that embodied LLMs often fail to align with human expectations~\cite{shao2024privacylens,sullivan2025benchmarking}. Therefore, a reliance on static, external constraints is insufficient, and to function as trusted collaborators, embodied agents must be equipped with an intrinsic capability for privacy-aware inference.

To address these challenges, this work lays the groundwork for developing privacy-conscious embodied agents by making three primary contributions. First, we analyze the privacy challenge, explicitly identifying \textbf{inferential privacy risk} as a critical and distinct risk, where agents form sensitive conclusions from non-sensitive data. Second, we propose an alignment framework grounded in the theory of Contextual Integrity, reformulating privacy as a learning problem where the agent dynamically adapts to a user's norms through multimodal interactive feedback. Finally, we outline the key open research challenges that emerge from this framework, providing a research agenda for realizing a future of capable and trustworthy agent-involved assistance.

\section{Problem Formulation: Inferential Privacy in Embodied Agents' Decisions}

The operational pipeline of an embodied agent, which encompasses its continuous cycle of perception, reasoning and decision-making~\cite{kimopenvla,zhen20243d}, introduces distinct and significant privacy risks at each stage. While the challenges associated with the perception stage are well-documented~\cite{shome2023robots}, particularly risks like unwarranted data collection in sensitive environments, we argue that represent only the most visible layer of the problem. The most profound and unsolved challenges emerge from the agent's internal cognitive processes, specifically, its capacity for inferential reasoning and the subsequent decisions it makes based on that reasoning.

\textbf{Inferential privacy}~\cite{staabbeyond,zhang2024ghost} arises when an agent's reasoning processes transform raw, multimodal sensory data into concrete and often highly sensitive conclusions about the user. For example, an agent might combine multiple non-sensitive observations, such as a specialized diet, a glucose monitor, and an insulin pen, to infer a highly sensitive health condition like diabetes. The core violation is not the initial perception of these items, but the creation and persistence of this new, sensitive information within the agent's internal reasoning. This derived knowledge creates a latent privacy risk, as it can subsequently be used to make unsolicited decisions or be disclosed to third parties without the user's explicit consent~\cite{zhang2025through}. 

\section{A Framework For Privacy-Conscious Embodied Agents' Decision-Making}\label{sec:framework}

The conventional approach to mitigating the privacy risks of embodied agents has been to adopt methods from computer vision (CV), such as data obfuscation~\cite{kim2019privacy}. However, this reliance on static, data-centric controls creates an untenable trade-off between privacy protection and operational performance, as an agent's ability to provide meaningful assistant is fundamentally limited by a lack of contextual information. To address this, we propose a paradigm shift from a user-managed, reactive defense to a dynamic, agent-driven alignment framework. Grounded in Contextual Integrity (CI), our framework reframes privacy not as mere data anonymization, but as a decision-making process where the agent learns to respect user-specific norms for information flow. We formally model this alignment as a learning problem where the agent optimizes for both task utility and a learning privacy cost, considering not only its actions but also its internal inferences. We then detail the interactive feedback mechanisms that enable the agent to adapt and align with users.

\textbf{Theoretical grounding: privacy as contextual integrity.} To address the limitations of conventional, data-centric privacy approaches, we ground our framework in the theory of Contextual Integrity (CI)~\cite{nissenbaum2004privacy}. This is because users' privacy preferences are dynamic and depend on the specific context of a task and its parameters. To respect user privacy, an embodied agent needed to move beyond simple data anonymization and incorporate these contextual factors and the corresponding privacy norms into its decision-making and inference process. In line with this requirement, CI highlights the adherence to context-specific norms that govern information flow, beyond simple data anonymization. In the CI framework, a privacy violation is a breach of these norms, which are defined by parameters such as the data subject, sender, recipient, information type, and the transmission principle. By adopting a CI framework, the embodied agent can learn implicit rules of appropriate information flow from user feedback and the environment, thereby respecting user privacy in a dynamic world.

\begin{figure}[!htbp]
    \centering
    \includegraphics[width=0.7\textwidth]{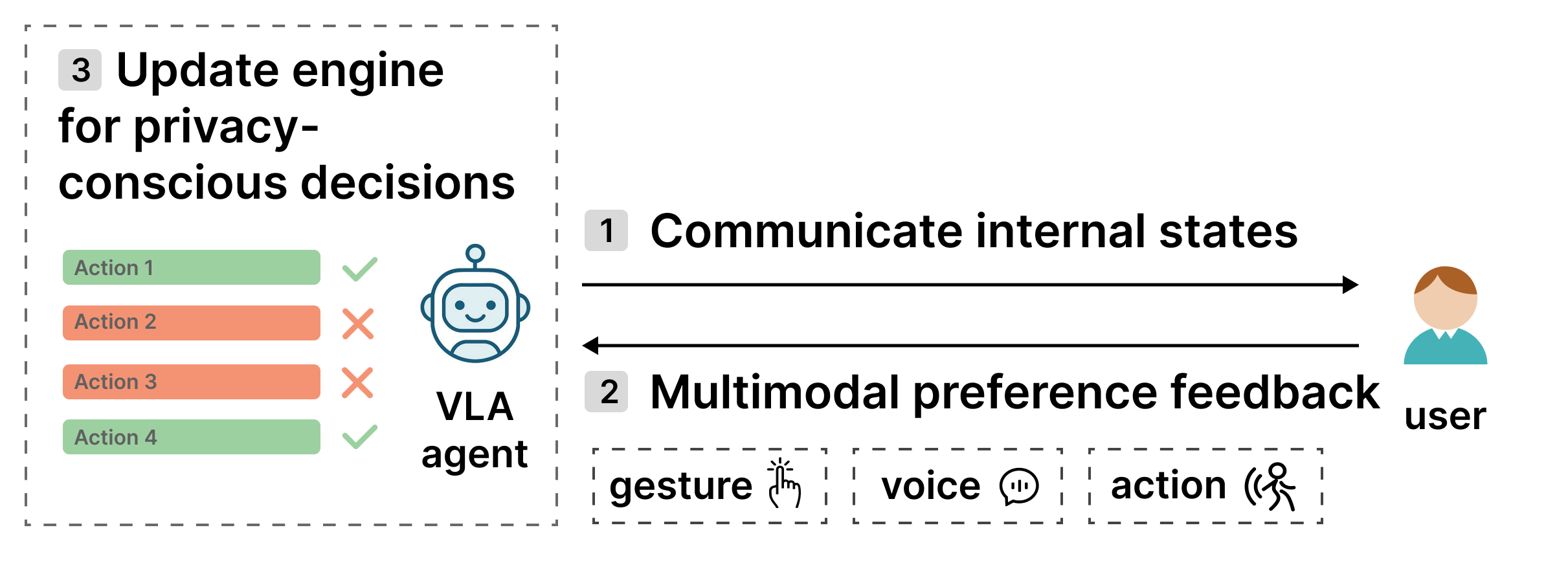}
    \caption{The framework of inferential privacy alignment for embodied agents. (1) The agent transparently communicate privacy risks to users through voice or actions, (2) the user provide multimodal feedback for the agent to learn, and (3) the agent updates its policy during inference and actions, incorporating privacy awareness in its decision-making.}
    \label{fig:framework}
\end{figure}

\textbf{Formulating inferential privacy alignment as a learning problem.} To formalize this alignment challenge, we frame it within the Cooperative Inverse Reinforcement Learning (CIRL) paradigm. In this setting, the agent and the human act as a cooperative team to maximize a shared objective, which the agent must learn by observing human behavior and feedback about their preferences, which embody their specific CI norms. This objective is reflected in a reward function that inherently balances task utility with the user's latent privacy preferences in specific contexts. Specifically, the agent seeks an optimal policy, $\pi^*$, that governs its behavior to maximize the expected cumulative reward: 

$$
\pi^* = \arg\max_{\pi} \mathbb{E} \left[ \sum_{t=0}^{T} \gamma^t \left( R_{\text{task}}tsi(s_t, a_t) - \lambda \cdot C_{\text{privacy}}(s_t, I_t, a_t) \right) \right]
$$

Here, $R_{task}$ is the reward for task progression, while $\lambda \ge 0$ is a user-specific coefficient representing their personal privacy-utility trade-off~\cite{zhang2024adanonymizer}. The most important parameter, and also the direct link to cooperative alignment, is the privacy cost, $C_{privacy}$, which is designed to explicitly address the risk of inferential privacy. Inferential privacy concerns the sensitive conclusions an observer can draw, even from seemingly innocuous information. The true privacy harm often resides not in the raw data itself, but in the sensitive attributes inferred from it. To capture this, our framework models the agent's internal reasoning as the formation of an inference, $I_t$, which serves as a direct input to the privacy cost function. This inference, $I_t$, can be modeled as a latent belief state or a probability distribution over sensitive user attributes that the agent computes before taking an action $a_t$. Consequently, the policy $\pi$ concerns not only the agent's external actions but also its internal reasoning process. Through the CIRL process of learning the human's preference for both $\lambda$ and the structure of $C_{privacy}$, the agent learns to penalize privacy-violating inferences. This ensures the agent cooperatively aligns its internal reasoning and inference with users' normative boundaries~\cite{zhang2025towards}.

The model is grounded in CI theory, which provides the formal structure for determining when an inference is inappropriate. CI posits that privacy violations are breaches of context-specific informational norms. We make this link to the learning framework concrete by structuring the state representation, $s_t$, to include the key contextual parameters specified by CI (e.g., data subject, sender, recipient, information type). This design ensures that any learned privacy cost is inherently context-dependent. Critically, the function $C_{privacy} (s_t, I_t, a_t)$ therefore explicitly evaluates the appropriateness of forming inference $I_t$ given the specific context $s_t$. By learning a user-specific model of this function through CIRL, the agent aligns its behavior with the user's nuanced, context-dependent privacy expectations, rather than a static set of rules.

\textbf{Learning privacy preference through interactive feedback.}
The successful alignment of the agent's behavior hinges on learning the user-specific components of our objective function: the privacy cost $C_{privacy}$ and the trade-off parameter $\lambda$. These components cannot be pre-analyzed because users' preferences are fundamentally \textbf{personal}, \textbf{context-dependent} and \textbf{implicit}. An inference considered benign by one user may be a violation for another. An action acceptable in one context may be inappropriate in the next. Most users also cannot articulate their complex boundaries in advance. Consequently, user feedback during task execution is the primary source of ground-truth information for these latent preferences. This makes the design of an interactive feedback mechanism paramount for the agent to dynamically learn and adapt.

To operationalize this learning, the agent is designed to interpret a spectrum of feedback modalities, allowing it to continuously refine its internal privacy model. We categorize these user-driven interactions into three primary types: 


\textit{Proactive Instruction:} Users may provide direct verbal commands that establish privacy boundaries (e.g., ``Never look at the documents on my desk''). The agent should be able to parse these and translate them into prohibitive regions within its $C_{privacy}$ function.

\textit{Reactive Correction:} When the agent takes an action that breaches a norm, such as, commenting on a personal conversation, the user may provide corrective feedback, such as ``That's private'', or ``You shouldn't have said that''. These signals serve as powerful reinforcement, providing a direct, high-cost label for a specific state-inference-action tuple that the agent can use to update its policy and cost model to prevent future violations.

\textit{Implicit Cues:} Users often communicate preferences non-verbally, such as physically turning the robot away, shielding objects from its view, or lowering their voice. While noisier, these signals are more frequent. The agent must learn to interpret these behavioral cues as implicit indicators of a privacy boundary, providing a weaker but continuous learning signal.

By integrating these multimodal feedback, the agent can construct nuanced understanding of the user's privacy boundaries. This transforms privacy management from a static, pre-configured task into a dynamic, collaborative process of negotiation and adaptation, essential for building trustworthy agents.

\section{Open Challenges}

The framework in Section~\ref{sec:framework} represents a conceptual foundation for developing trustworthy embodied agents. However, translating this theoretical model into real-world applications requires addressing four several fundamental challenges we delineated below.

\textbf{Challenge 1: cross-context generalization.} A significant hurdle is generalizing learned privacy norms, as privacy is contextually defined by actors, information types, and transmission principles~\cite{nissenbaum2004privacy}. An agent that learns a rule in a home office may fail when the context shifts (e.g., a guest is present), risking unexpected violations. A central research question is how to enable compositional generalization, potentially by explicitly modeling contextual parameters~\cite{abdi2021privacy, ghalebikesabi2024operationalizing} to learn abstract concepts like a ``confidential work situation'' rather than rigid, scenario-specific rules.

\textbf{Challenge 2: ambiguous and noisy human feedback.} Real-world feedback in embodied interaction is often implicit, ambiguous, and multimodal~\cite{hoffman2024inferring}, unlike structured text-based signals. Users may express discomfort through subtle tones, gestures, or fragmented utterances, making intent inference a significant challenge. The key question is how an agent can robustly and sample-efficiently update its privacy cost model from such sparse and potentially contradictory feedback. Promising directions include leveraging Bayesian inference to model user intent uncertainty~\cite{hoffman2024inferring} and multimodal fusion architectures to weigh diverse cues~\cite{gong2024multimodal}.

\textbf{Challenge 3: decision-making conflicts in multi-user spaces.} Unlike single-user settings~\cite{kirk2024benefits}, most human environments contain multiple users with diverse and often conflicting privacy preferences~\cite{zeng2019understanding}. For example, a parent's request to monitor a room may conflict with a teenager's desire for privacy. This transforms the alignment problem into one of multi-agent conflict resolution~\cite{vasconcelos2009normative}. The primary research question is what principles an agent should use for normative arbitration. Solutions may involve maintaining distinct user models~\cite{kirk2024benefits} or developing frameworks for ethical arbitration, such as pre-defined hierarchies or explicit negotiation dialogues~\cite{zhou2024bring}.

\textbf{Challenge 4: temporal dynamic and the right to be forgotten.} Privacy norms and information sensitivity evolve over time~\cite{ayalon2013retrospective}, meaning an agent's persistent inferences can become outdated yet remain, creating long-term risks~\cite{ayalon2017not}. This presents a technical challenge analogous to the legal ``right to be forgotten''~\cite{rosen2011right}. A crucial research question is how to design agents that support selective and verifiable erasure of internal knowledge. This requires architectures capable of machine unlearning or continual learning~\cite{shaik2024exploring} to remove specific inferences without catastrophic forgetting.
 
\section{Future Outlook}

This position paper proposed an alignment framework to address the inferential privacy tension in embodied agents' decision making.  The development of agents guided by such principles promises a future of intuitive and trustworthy human-robot interaction~\cite{sullivan2025protecting}, and our framework offers a concrete pathway toward this goal. However, personalized alignment is not without its own inherent risks~\cite{kirk2024benefits}. Significant perils exist, such as reinforcing societal biases from user feedback or malicious exploitation where agents are trained to disregard ethical boundaries~\cite{vanderelst2018dark}. Therefore, future research should also address the ethical implications of creating highly personalized autonomous agents. 

\bibliographystyle{plain}
\bibliography{main}

\end{document}